%%%%%%%%%%%%%%%%%%%%%%%%%%%%%%%%%%%%%%%%%%%%%%%%%%%%%%%%%%%%%%%%%%%%%%%%%%%%%%
\magnification\magstep 1                                                      
\baselineskip=0,59 true cm
\vsize=21 true cm
\topinsert\vskip 2 true cm
\endinsert
{\centerline{\bf{THE GROUP $\Gamma(2)$ AND THE FRACTIONAL QUANTUM 
HALL EFFECT}}}
\vskip 2 true cm
{\centerline{\bf{Yvon Georgelin, Jean-Christophe Wallet}}}
\vskip 1 true cm
{\centerline{$^a$ Division de Physique Th\'eorique{\footnote
\dag{\sevenrm{Unit\'e de Recherche des Universit\'es Paris 11 et Paris 6
associ\'ee au CNRS}}}, Institut de Physique
Nucl\'eaire}}
{\centerline{F-91406 ORSAY Cedex, France}}
\vskip 2 true cm
{\bf{Abstract:}} \par
We analyze the action of the inhomogeneous modular group $\Gamma(2)$ on the
three cusps of its principal fundamental domain in the Poincar\'e half plane.
From this, we obtain an exhaustive classification of the fractional quantum
Hall numbers. This classification is somehow similar
to the one given by Jain. We also present some resulting remarks 
concerning direct 
phase transitions between the different quantum Hall states.
\vskip 3 true cm
{\noindent{IPN}}O-TH-9618 (March 1996)\hfill\break
\vfill\eject
\def\init{\tabskip 0pt\offinterlineskip}
\def\crr{\cr\noalign{\hrule}}
{\bf{I) Introduction}}\hfill\break
\vskip 0,5 true cm
{\noindent{Since the}} discovery of the two dimensional quantized
Hall conductivity by [1]
for the integer plateaus and by [2] for the fractional ones, the quantum Hall 
effect has been a constant and intensive field of theoretical and
experimental investigations.\par
As far as the theoretical viewpoint is concerned, pionneering contributions
by [3a-c] have related the basic features of the hierarchy of the Hall 
plateaus with the properties of a two dimensional incompressible fluid with
fractional charges collective excitations [3a]. This has been 
followed by a large
number of related works, dealing with condensed matter theory [4], 
quantum field theory [5], and mathematical physics [6].\par
The purpose of this paper is to set up a link between the similarity
transformations as introduced by Jain [7] in the 
hierarchical model of 
[3a-c] and
some basic operations of the group $\Gamma(2)$, which is known in the 
mathematical literature as the level 2--principal congruence subgroup of 
the inhomogeneous modular group $\Gamma(1)$ [8]. It will be shown 
that the similarity
transformations of Jain [7] (see also [10]) may well be described 
as a family of 
specific transformations of $\Gamma(2)$ acting on the 
three cusps $[i\infty]$, $[0]$ and
$[1\simeq -1]$ (that is, the well known vertices of one of its fundamental
domains [9]). In particular, the whole structure
of the integer and fractional plateaus can be entirely recovered from this
family of transformations, independantly of any microscopic 
models. Furthermore, the results obtained from this arithmetical
construction may provide a new insight on recently proposed Hall phases
diagrams [10-12], as it will be briefly discussed.\par
\vskip 0,5 true cm
{\bf{II) $\Gamma(2)$ and the quantum Hall effect. }}\hfill\break
{\bf{IIa -- Basic properties}}\par
{\noindent{Let ${\cal{P}}$ be }}the upper complex plane and $z$ a complex 
coordinate ($Im\ z> 0$) on ${\cal{P}}$. We recall that the transformations $G$
acting on ${\cal{P}}$ and
pertaining to the inhomogeneous modular group $\Gamma(1)$ are defined by
$$G(z)={{az+b}\over{cz+d}};\ a,b,c,d \in Z;\ ad-bc=1 ({\hbox{unimodularity
condition}}) \eqno(1).$$
The group $\Gamma(1)$ leaves also invariant the real numbers and the set of
rational numbers.\par
Up to now, two subgroups of $\Gamma(1)$ 
underlying the main models for the
Hall effect have essentially been considered. The first one, $\Gamma_\theta$
($\equiv\Gamma_S(2)$), appears to be the natural symmetry group of the 
Landau problem on a torus; it is generated by the following transformations
on ${\cal{P}}$:
$$T^2(\tau)=\tau+2;\ S(\tau)=-{{1}\over{\tau}}  \eqno(2a;b).$$
Here, the complex parameter $\tau$ characterizes the considered torus. As 
clearly demonstrated in [13], the restriction to the even translations is
required in order to recover the exact periodic conditions on the wave
functions, up to an allowed gauge transformation. This subgroup appears also in
the construction of the many body Landau states on the torus [14]
in terms of the
Coulomb gaz vertex operators [15]. One of the reasons why $\Gamma_\theta$ can
be hardly retained as a good classifying symmetry for the physical Hall 
system is that it has only two equivalence classes among the real 
fractions $p/q$ (namely $pq\in 2Z$ and $pq\in 2Z+1$), which apparently do not
correspond to any observed feature in the quantum Hall hierarchy.\par
The second relevant subgroup of $\Gamma(1)$ is related to the similarity
transformations of Jain used in the hierarchical model. In this 
framework, two fundamental exchange symmetries acting on the fractional Hall
numbers $\nu=p/q$ have been singled out [7]. They are defined by
$$\nu \iff \nu +1;\ \nu \iff {{\nu}\over{2\nu+1}}  \eqno(3a;b).$$
Physically, (3a) corresponds to the Landau level addition transformations
whereas (3b) represents flux attachement transformations (see [7] and [10]).
Another exchange symmetry is also present at least for the first Landau
level; it is given by $\nu \iff 1-\nu$ and represents a
particle-hole symmetry. It is an
important remark that (3a) and (3b) generate the subgroup $\Gamma_0(2)$($\equiv
\Gamma_T(2)$) of $\Gamma(1)$. Notice also that $\Gamma_0(2)$ 
has been proposed in [10-12] as the best
candidate for a dynamical flow symmetry group generated in renormalization
group equations for the complex Hall conductivity. Its nice feature is that it
respects the parity of the denominators when acting on the fractions $p/q$.
However, the way how it could take into account finer structures as well as
the particle-hole similarity is not so clear.\par
One of the reasons leading us to look for other subgroups 
of $\Gamma(1)$ sufficient to generate the Hall fractional numbers 
hierarchy comes from a
comparison between the experimental situation for the Hall system and the
(many body) Landau problem on the torus. Indeed, it has been
underlined in [16] that in the true macroscopic 
Hall experiment, the sample with the threads attached to it
can be viewed as a topological subsystem
of the (many body) Landau problem on the torus (see figure 1). This is to be
contrasted with the mathematical property that $\Gamma_0(2)$ 
and $\Gamma_\theta$
bear no inclusion relation the one with respect to the other but instead are
conjugate groups in $\Gamma(1)$. Therefore, if one considers roughly 
the Hall system as a kind of
subsystem of the Landau problem on the torus, it is quite natural to look for a
subgroup of $\Gamma(1)$, included both in $\Gamma_\theta$ and
$\Gamma_0(2)$, and which can generate the quantum Hall number hierarchy.\par
We find that the largest group sharing the above properties
is the inhomogeneous principal congruence group
at level 2, usually denoted by $\Gamma(2)$ in the mathematical literature. Its
action on ${\cal{P}}$ is given by
$$G(z)={{az+b}\over{cz+d}};\  
ad-bc=1\ {\hbox{(unimodularity condition)}};\ a,d\ 
{\hbox{odd}};\ b,c\ {\hbox{even}}  \eqno(4).$$
It is a free group generated by
$$T^2(z)=z+2;\ \Sigma(z)\equiv ST^{-2}S(z)={{z}\over{2z+1}}  \eqno(5),$$
and the following inclusion relation holds:
$$\Gamma(2)\subset(\Gamma_0(2),\Gamma_\theta)\subset\Gamma(1)  \eqno(6).$$
The main properties of $\Gamma(2)$ can be found e.g in [8], [9]. For the present
purpose, it is crucial to recall that its principal
fundamental domains ${\cal{D}}$, is
obtained by identifying the frontiers as indicated on figure (2). Moreover,
${\cal{D}}$ has three cusps denoted by
$[i\infty]$, $[0]$ and $[1\simeq-1]$ which are identified with the three
integers $\infty$, $0$ and $1$. \par
Corresponding to this, the action of $\Gamma(2)$ on real rational 
fractions generates also three well
separated orbits on the rational numbers
$$\{ {{2k}\over{2m+1}}\};\ \{ {{2k+1}\over{2m+1}}\};\ 
\{ {{2k+1}\over{2m}}\}\cup\{\infty\};\ 
  k,m\in Z   \eqno(7),$$
suggesting that it could be of some relevance for the quantum Hall hierarchy
[12]. Actually,
we claim that the action of $\Gamma(2)$ on the cusps allows one to describe
a more detailed structure in this hierarchy. Our proposal is the 
following: The action of $\Gamma(2)$ on the three 
cusps $[i\infty]$, $[0]$ and $[1\simeq-1]$ generates natural families of
transformations in one-to-one correspondance with families of the Jain
hierarchy, which permits one to obtain an exhaustive classification of all 
the fractional Hall numbers observed up to now. It could give also some new
light on the structure of the phase diagram for the quantum Hall effect.\par
\vskip 0,5 true cm
{\bf{IIb -- The construction.}}\par
{\noindent{Let us }}choose the following parametrization for $G\in\Gamma(2)$
$$G(z)={{(2s+1)z+2n}\over{2rz+(2k+1)}}\ \ \ k,n,r,s\in Z  \eqno(8).$$
The unimodularity condition imposes
$$(2s+1)(2k+1)-4rn=1  \eqno(9).$$
Let us now select $\{G^\lambda\}$, a family of transformations 
in $\Gamma(2)$ parametrized by $\lambda$, such that each $G^\lambda$ sends the
cusp $[i\infty]$ onto a given irreducible {\footnote\dag{\sevenrm{
The fraction must be irreducible because of the unimodularity condition}}}
fraction with even denominator, 
$\lambda={{2s+1}\over{2r}}$ (so that $lim_{\vert
z\vert\to\infty}G^\lambda(z)=\lambda$). Then, it can be easily realized that
the image by the $G^\lambda$'s of the two 
other cusps $[0]$ and $[1\simeq -1]$ of
${\cal{D}}$ generates precisely a Jain hierarchy.\par
Indeed, let us pick for example $\lambda=1/2$; then any transformation
$G^{1/2}$ which sends the cusp $[i\infty]$ onto the
irreducible fraction $\lambda={{1}\over{2}}$ (implying that 
$s=0$ and $r=1$) could be parametrized, using (8) as 
$$G^{1/2}(z)={{z+2n}\over{2z+2k+1}},\ k,n\in Z  \eqno(10),$$
together with the unimodularity condition which implies $2k=4n$; then, 
(10) becomes
$$G^{1/2}_n(z)={{z+2n}\over{2z+4n+1}}  \eqno(11).$$
Now, the images by $G^{1/2}_n$ of the cusps $[0]$ and $[1\simeq-1]$ are given by
$$G^{1/2}_n(0)={{2n}\over{4n+1}};\ G^{1/2}_n(1)={{2n+1}\over{4n+3}}  
\eqno(12),$$
corresponding to the families of fractional levels collected in table 1. There,
one easily identifies the Hall fractions located on each side of the
$\lambda=1/2$ fraction; this includes the numbers 0 and 1, the integer 
Hall numbers which border this family.\par
Starting from $\lambda=3/4$, a similar construction gives 
rise to the Hall numbers
family located on each side of the $\lambda=3/4$ fraction, including 
the border numbers 1 and $2/3$. Indeed, one must have
$$G^{3/4}(i\infty)={{3}\over{4}}  \eqno(13).$$
This, combined with (8) and (9) (implying $r=2$ and $s=1$) yields
$$G^{3/4}_n(z)={{3z+2n}\over{4z+(2k+1)}},\ n,k\in Z;\ 3k=4n-1\ 
{\hbox{(unimodularity)}}  \eqno(14a;b).$$
From this later relations, the images of $[0]$ and $[1\simeq-1]$ under
the $G^{3/4}_n$'s are easily computed. The results, collected in table 2, are 
nothing but the Hall numbers family located on each side of the $\lambda=3/4$
fraction, including the borders 1 and $2/3$.\par
The construction proceeds in the same way , starting also from any 
irreducible even denominator fraction larger than 1. Pick for 
example $\lambda=3/2$. One has
$$G^{3/2}_n(z)={{3z+2n}\over{2z+(2k+1)}},\ n,k\in Z;\ 3k=2n-1\ 
{\hbox{(unimodularity)}}  \eqno(15a;b).$$
The images of $[0]$ and $[1\simeq-1]$  by $G^{3/2}$ are collected in table 3,
which reproduces the Jain hierarchy symmetrical about
$\lambda=3/2$, including the borders 1 and
2. \par
The results for $\lambda=1/4$ are also collected in table 4.\par
\vskip 0,5 true cm
{\bf{III) Discussion and conclusion. }}\hfill\break
Summarizing, we claim that any Jain hierarchy symmetrical about 
any fraction $\lambda$ with even denominator is the image of the cusps $[0]$
and $[1\simeq-1]$ of ${\cal{D}}$ by the family of transformations
$G^\lambda\in\Gamma(2)$ sending the cusp $[i\infty]$ to $\lambda$. The fraction
$\lambda$ belongs to the orbit $\{(2m+1)/2r\}\cup\{\infty\}$. \par
The complete sequences of Hall fractionals is of course not still
experimentally determined but the hierarchical families obtained from our
construction are in complete agreement with what is up to now 
experimentally confirmed [17].\par
It has been recently argued that it can be associated to the Hall system a
two-dimensional (i.e magnetic field -- disorder variable) phase diagram [10-12] 
where the even denominator fractions label Hall metallic states whereas the odd
denominator fractions as well as the non zero integers label Hall liquid
states, the number zero corresponding to a unique Hall insulator state. In that
framework, each Hall metallic state appears to be surrounded by a defined
family of Hall liquid states, the Hall insulator state being "in contact" with
a restricted number of other states [10].\par
Our construction agrees globally with this picture. Here too, a 
central role is played
by the (metallic) even denominator fractions, each of them generating two
symmetric families of odd denominator and integer Hall liquid states
(corresponding to the action of the $G^\lambda$'s on the 
cusps $[0]$ and $[1]$). In
each of these families labeled by $\lambda$, it could eventually appears (see
tables)
the Hall insulator $\nu=0$ (for exemple, this happens for $\lambda=1/2,\
1/4$ but not for $\lambda=3/4,\ 3/2$). We suspect in that case the occurence of 
a {\it{direct}} insulator/metal phase 
transition. Furthermore, according to our
description ($\nu=0$ appears only in $G^\lambda(0)$), a direct transition
between the insulator and the nearest liquid state appearing in $G^\lambda(1)$
is also probably allowed (see tables). It is amazing to observe that, if we
accept such a "neighboring principle" for possible direct Hall phase
transitions between the different entries of our tables, we actually predict
a direct phase transition between the insulator and the liquid states
$2/3,2/5,2/7,2/9$ (see table 1 and 4). In this respect, our classification
is closer to the one proposed in [18] rather than the one given by [10-11].\par
The complete description of the Hall phase diagram is still subject to some
controversy. Nevertheless, our present mathematical (arithmetical)
construction fits quite well with the existing 
datas [17,18]. \par                     
Let us add a comment concerning the interpretation of the particle-hole
similarity [7] $\nu^\prime+\nu=1$ within our construction. Clearly, this last
relation can be satisfied only for $0\le\nu(\nu^\prime)\le 1$ (i.e. inside the
first Landau level). The way we constructed the hierarchies from 
the $G^\lambda$'s indicates we should have $0<\lambda<1$. Now, it is known
in arithmetics that given a positive odd number $2m+1$, the set of fractions
with that given denominator obtained from our present construction is 
the same as the set of fractions with denominators equal to $2m+1$ belonging 
to the order $2m+1$ Farey sequence $F_{2m+1}$ [19]. It is a nice property 
of Farey sequences that the sum of two symmetric fractions about $\lambda=1/2$
is equal to 1. This is identical to $\nu^\prime+\nu=1$. We
can translate this in terms of our $G^\lambda$'s: the particle-hole similarity
inside the first Landau level corresponds to the relation
$$G^\lambda(0)+G^{\lambda^\prime}(1)=1  \eqno(16),$$
which holds only when
$$\lambda+\lambda^\prime=1\ \ \lambda,\lambda^\prime>0, \ 
\lambda,\lambda^\prime<1    \eqno(17).$$
The solution of these equations gives any pairs of fractions related by a
particle-hole similarity. For example, pick $\lambda=\lambda^\prime=1/2$, then
$G_n^{1/2}(0)+G_{n^\prime}^{1/2}(1)=1$ is 
verified provided $n+n^\prime=-1$.\par
One mathematical remark is now in order. Among all the inhomogenous
modular subgroups of
$\Gamma(1)$, the group $\Gamma(2)$ is the only one possessing the principal
fundamental domain ${\cal{D}}$ with only three cusps $[0]$, $[1\simeq -1]$ and
$[i\infty]$. In this sense, our construction is unique.\par
The physical  interpretation (if any) of the 
cusps is obscure to us at the present time. The same remark could be
made concerning 
the "shift operators" $G^\lambda$'s.  Nevertheless, as a final conclusion, we
recall that microscopic models
proposed recently [20] to describe the quantum Hall dynamics exhibit
an approximate
$\Gamma(1)$ invariance from which some physical usefull observables (for
instance the height of each longitudinal conductivity peaks) could in principle
be determined. We suggest here that $\Gamma(2)$ could in fact be an alternative
candidate to $\Gamma(1)$ as a dynamical group to be exploited 
as a starting point to built future microscopic models.\par
\vfill\eject
\topinsert\vskip 2 true cm
\endinsert
{\bf{REFERENCES}}\par
\vskip 2 true cm
\item {[1]:} K. von Klitzing, G. Dorda and M. Pepper, Phys. Rev. Lett. 45 (1980)
494.
\item {[2]:} D. C. Tsui, H. L. St\"ormer and A. C. Gossard, Phys. Rev. Lett. 48
(1982) 1559.
\item {[3a]:} R. B. Laughlin, Phys. Rev. Lett. 50 (1983) 1395.
\item {[3b]:} F. D. M. Haldane, Phys. Rev. Lett. 51 (1983) 605.
\item {[3c]:} B. I. Halperin, Phys. Rev. Lett. 52 (1984) 1583.
\item {[4]:} For a review, see, e.g. Quantum Hall effect, by M. Stone, World
Scientific (Singapore) 1992.
\item {[5]:} We have in mind models related to Chern-Simons theory, Coulomb gaz
approach and rational conformal field theories; see [4] and references therein.
\item {[6]:} A. Connes, in ``G\'eom\'etrie non commutative, InterEditions,
Paris (1990); J. Xia, Comm. Math. Phys. 119 (1988) 29.
\item {[7]:} J. K. Jain, Phys. Rev. Lett. 63 (1989) 199; Phys. Rev. B41 (1990)
7653.
\item {[8]:} See in D. Mumford, Tata Lectures on Theta (vols. I, II, III),
Birkh\"auser, Boston, Basel, Stuttgart (1983).
\item {[9]:} See in B. Schoeneberg, Elliptic Modular Functions, an
introduction, Springer-Verlag, Berlin, Heidelberg (1974).
\item {[10]:} S. Kivelson, Dung-Hai Lee and Shou-Cheng Zhang, Phys. Rev. B46
(1992) 2223.
\item {[11]:} B. I. Halperin, P. A. Lee and N. Read, Phys. Rev. B47 (1993)
7312.
\item {[12]:} C. A. L\"utken and G. G. Ross, Phys. Rev. B48 (1993) 2500; Phys.
Rev. B48 (1993) 11837; C. A. L\"utken, Nucl. Phys. B396 (1993) 670.
\item {[13]:} P. L\'evay, J. Math. Phys. 36 (1995) 2792.
\item {[14]:} F. D. M. Haldane and E. H. Rezayi, Phys. Rev. B31 (1985) 2529.
\item {[15]:} G. Christofano, G. Maiella, R. Musto and F. Nicomedi, Phys. Lett.
B262 (1991) 88.
\item {[16]:} J. E. Avron and R. Seiler, Phys. Rev. Lett. 54 (1985) 259; see
also Q. Niu and D. J. Thouless, Phys. Rev. B35 (1987) 2188.
\item {[17]:} see e.g. R. G. Clark et al., Phys. Rev. Lett. 60 (1988) 1747; 
J. P. Eisenstein, H. L. Stormer, L. N. Pfeiffer and K.
W. West, Phys. Rev. B41 (1990) 7910; H. W. Jiang et al., Phys. Rev. B44 (1991)
8107; L. W. Engel et al., Phys. Rev. B47 (1992) 3418; M. B. Santos et al.,
Phys. Rev. B46 (1992) 13639; R. R. Du et al., Phys. Rev. Lett. 70 (1992) 
2944; H. C. Manoharan and M. Shayegan, Phys. Rev. B50 (1994) 17662.
\item {[18]:} S. V. Kravchenko, W. Mason, J. E. Furneaux and V. M. Pudalov to
be published in Phys. Rev. Lett.
\item {[19]:} see T. M. Apostol, Modular functions and Dirichlet series in
number theory, Springer-Verlag, New-York, Berlin, Heidelberg (1976).
\item {[20]:} E. Fradkin and S. Kivelson, preprint University of
Urbana-Champaign (1996).
\vfill\eject
\topinsert\vskip 2 true cm
\endinsert
{\bf{FIGURE CAPTIONS}}\par
\vskip 2 true cm
{\bf{Figure 1}}: A Hall sample "sitting" on a torus with the thickness  
of the threads largely exagerated; the magnetic field is constant and    
perpendicular to the sample.\par
\vskip 3 true cm
{\bf{Figure 2}}: The principal fundamental domain of $\Gamma(2)$         
in the Poincar\'e plane showing the cusps $[0]$, $[1\simeq-1]$ and       
$[i\infty]$. The identification operations $T^2$ and $\Sigma$ are        
indicated.\par

\vfill\eject
\hsize=20 true cm
{\bf{TABLE 1: $\lambda=1/2$; $(k=2n$)}}\par
$$\vbox{\init\halign to 19 true cm{
\strut#&\vrule#\tabskip=1em plus 2em&
\hfil$#$\hfil&
\vrule$\,$\vrule#&
\hfil$#$\hfil&
\vrule#&
\hfil$#$\hfil&
\vrule#&
\hfil$#$\hfil&
\vrule#&
\hfil$#$\hfil&
\vrule#&
\hfil$#$\hfil&
\vrule#&
\hfil$#$\hfil&
\vrule#&
\hfil$#$\hfil&
\vrule#&
\hfil$#$\hfil&
\vrule#&
\hfil$#$&
\vrule#\tabskip 0pt\crr
&&n&&...&&-3&&-2&&-1&&0&&1&&2&&3&&...&\crr
&&k&&...&&-6&&-4&&-2&&0&&2&&4&&6&&...&\crr
&&G^{1/2}_n(0)&&...&&{{6}\over{11}}&&{{4}\over{7}}&&{{2}\over{3}}&&
0&&{{2}\over{5}}&&{{4}\over{9}}&&{{6}\over{13}}
&&...&\crr
&&G^{1/2}_n(1)&&...&&{{5}\over{9}} &&{{3}\over{5}} 
&&1 &&{{1}\over{3}}&&{{3}\over{7}}
&&{{5}\over{11}}&&
{{7}\over{15}}&&...&\crr
}}$$
\vskip 1 true cm
{\bf{TABLE 2: $\lambda=3/4$; $(3k=4n-1$)}}\par
$$\vbox{\init\halign to 19 true cm{
\strut#&\vrule#\tabskip=1em plus 2em&
\hfil$#$\hfil&
\vrule$\,$\vrule#&
\hfil$#$\hfil&
\vrule#&
\hfil$#$\hfil&
\vrule#&
\hfil$#$\hfil&
\vrule#&
\hfil$#$\hfil&
\vrule#&
\hfil$#$\hfil&
\vrule#&
\hfil$#$\hfil&
\vrule#&
\hfil$#$\hfil&
\vrule#&
\hfil$#$\hfil&
\vrule#&
\hfil$#$&
\vrule#\tabskip 0pt\crr
&&n&&...&&-8&&-5&&-2&&1&&4&&7&&...&&...&\crr
&&k&&...&&-11&&-7&&-3&&1&&5&&9&&...&&...&\crr
&&G^{3/4}_n(0)&&...&&{{16}\over{21}}&&{{10}\over{13}}&&{{4}\over{5}}&&
{{2}\over{3}}&&{{8}\over{11}}&&{{14}\over{19}}&&...
&&...&\crr
&&G^{3/4}_n(1)&&...&&{{13}\over{17}}&&{{7}\over{9}}&&1 &&{{5}\over{7}}
&&{{11}\over{15}}
&&{{17}\over{23}}&&
...&&...&\crr
}}$$
\vskip 1 true cm
{\bf{TABLE 3: $\lambda=3/2$; $(3k=2n-1)$}}\par
$$\vbox{\init\halign to 19 true cm{
\strut#&\vrule#\tabskip=1em plus 2em&
\hfil$#$\hfil&
\vrule$\,$\vrule#&
\hfil$#$\hfil&
\vrule#&
\hfil$#$\hfil&
\vrule#&
\hfil$#$\hfil&
\vrule#&
\hfil$#$\hfil&
\vrule#&
\hfil$#$\hfil&
\vrule#&
\hfil$#$\hfil&
\vrule#&
\hfil$#$\hfil&
\vrule#&
\hfil$#$\hfil&
\vrule#&
\hfil$#$&
\vrule#\tabskip 0pt\crr
&&n&&...&&-7&&-4&&-1&&2&&5&&8&&...&&...&\crr
&&k&&...&&-5&&-3&&-1&&1&&3&&5&&...&&...&\crr
&&G^{3/2}_n(0)&&...&&{{14}\over{9}}&&{{8}\over{5}}&&2&&
{{4}\over{3}}&&{{10}\over{7}}&&{{16}\over{11}}&&...
&&...&\crr
&&G^{3/2}_n(1)&&...&&{{11}\over{7}}&&{{5}\over{3}} &&1 &&{{7}\over{5}}
&&{{13}\over{9}}
&&{{19}\over{13}}&&
...&&...&\crr
}}$$
\vskip 1 true cm

{\bf{TABLE 4: $\lambda=1/4$; $(k=4n)$}}\par
$$\vbox{\init\halign to 19 true cm{
\strut#&\vrule#\tabskip=1em plus 2em&
\hfil$#$\hfil&
\vrule$\,$\vrule#&
\hfil$#$\hfil&
\vrule#&
\hfil$#$\hfil&
\vrule#&
\hfil$#$\hfil&
\vrule#&
\hfil$#$\hfil&
\vrule#&
\hfil$#$\hfil&
\vrule#&
\hfil$#$\hfil&
\vrule#&
\hfil$#$\hfil&
\vrule#&
\hfil$#$\hfil&
\vrule#&
\hfil$#$&
\vrule#\tabskip 0pt\crr
&&n&&...&&-3&&-2&&-1&&0&&1&&2&&...&&...&\crr
&&k&&...&&-12&&-8&&-4&&0&&4&&8&&...&&...&\crr
&&G^{1/4}_n(0)&&... &&{{6}\over{23}} &&{{4}\over{15}}&&{{2}\over{7}}&&
0&&{{2}\over{9}}&&{{4}\over{17}}&&...
&&...&\crr
&&G^{1/4}_n(1)&&...&&{{5}\over{19}} 
&&{{3}\over{11}} &&{{1}\over{3}} &&{{1}\over{5}}
&&{{3}\over{13}}
&&{{5}\over{21}}&&
...&&...&\crr
}}$$

\end